\documentclass[aps,prl,twocolumn,groupedaddress,amsmath,amssymb]{revtex4-1}
\usepackage{graphicx}  
\usepackage{dcolumn}   
\usepackage{bm}        
\usepackage{verbatim}   
\usepackage{amssymb}
\usepackage{amsfonts}
\usepackage{amsmath}
\usepackage{graphicx}
\usepackage{bm}
\usepackage{subfig,tikz}
\usepackage{wrapfig}
\usepackage{xspace}
\usepackage{enumerate}
\newtheorem{theorem}{Theorem}

\newcommand{\R}{\mathbb R}

\begin{document}
\title{Mass-Angular Momentum Inequality For Black Ring Spacetimes}

\author{Aghil Alaee}
\address{Department of Mathematical and Statistical Sciences, University of Alberta, Edmonton, AB T6G 2G1, Canada, The Fields Institute for Research in Mathematical Sciences, Toronto, ON M5T 3J1, Canada}
\email{khangha@ualberta.ca}
\author{Marcus Khuri}
\address{Department of Mathematics, Stony Brook University, Stony Brook, NY 11794, USA}
\email{khuri@math.sunysb.edu}
\author{Hari Kunduri}
\address{Department of Mathematics and Statistics, Memorial University of Newfoundland, St John's NL A1C 4P5, Canada}
\email{hkkunduri@mun.ca}


\begin{abstract}  \noindent
The inequality $m^3\geq \frac{27\pi}{4} |\mathcal{J}_{2}|\left|\mathcal{J}_{1}-\mathcal{J}_{2}\right|$ relating total mass and angular momenta, is established for (possibly dynamical) spacetimes admitting black holes of ring ($S^1\times S^2$) topology.
This inequality is shown to be sharp in the sense that it is saturated precisely for the extreme
Pomeransky-Sen'kov black ring solutions. The physical significance of this inequality and its relation to new evidence of black ring instability, as well as the standard picture of gravitational collapse, are discussed.
\end{abstract}
\maketitle





The standard picture of gravitational collapse rests on two conjectures. Namely, \textit{weak cosmic censorship} (WCC) asserts that collapse always results in a black hole, and the \textit{final state conjecture} (FSC) contends that spacetime must settle down to a stationary
(electro-)vacuum final state. In four dimensions the no hair theorem \cite{chrusciel2012stationary} implies that the final state must then be a Kerr black hole (electromagnetic contributions will be ignored for simplicity).
Although angular momentum may be radiated away with gravitational waves, if the spacetime is
axisymmetric then total angular momentum is conserved so that the angular momentum of an initial
state $\mathcal{J}$ agrees with that of the final state $\mathcal{J}_0$. In addition, since gravitational radiation also carries away positive energy the mass of an initial state must be larger than that of the final state, $m\geq m_0$. These observations then yield the following inequality
between mass and angular momentum for any initial state
\begin{equation}\label{a1}
m^2\geq|\mathcal{J}|,
\end{equation}
since it is also satisfied by the Kerr final state having mass $m_0$ and angular momentum $\mathcal{J}_0$.

In higher dimensions $D>4$ all known stationary vacuum black hole spacetimes admit multiple rotational symmetries, and are thus multi-axisymmetric. A typical symmetry group is $U(1)^{D-3}$, and the only dimension greater than four for which this amount of symmetry is compatible with an asymptotically flat structure, necessary for the ADM mass, is five. The generalization of Kerr to $D=5$ is given by the Myers-Perry solution \cite{Myers1986}, which has a spherical $S^3$ black hole topology and a $U(1)^2$ symmetry with two associated angular momenta $\mathcal{J}_{i}$, $i=1,2$. A similar derivation as above, with the role of Kerr played by the Myers-Perry black hole, shows that any initial state of a 5-dimensional black hole spacetime with $S^3$ horizon topology should satisfy
\begin{equation}\label{a2}
m^3\geq \frac{27\pi}{32}\left(|\mathcal{J}_1|+|\mathcal{J}_2|\right)^2.
\end{equation}
Because the derivations of \eqref{a1} and \eqref{a2} rely so heavily on WCC and FSC, any violation of these mass-angular momentum inequalities would provide a counterexample to the
standard picture of gravitational collapse in their respective dimensions. On the other hand a rigorous verification of \eqref{a1} and \eqref{a2} lends credence to the standard picture, since there does not seem to be an alternate explanation for why such nontrivial inequalities should hold other than the arguments based on WCC and FSC. In fact, both inequalities have been proven
\cite{Dain2008, alaee2016proof}.

An important special case of FSC is the question of stability for stationary vacuum black holes.
Kerr is known to be linearly stable \cite{Finster:2016tky,Whiting}, and although a complete proof has not yet been given all evidence strongly supports the conclusion that it is nonlinearly stable as well \cite{Dafermos:2010hb,Dafermos:2014cua,Dafermos:2014jwa}. Although ultraspinning instabilities have been observed \cite{Dias:2009iu,Dias:2010eu,Reall0} in $D>5$ Myers-Perry black holes, in $D=5$ the Myers-Perry solution is expected to be linearly stable \cite{Reall}. Thus it is not a surprise that \eqref{a1} and \eqref{a2} have been confirmed, as this is consistent with the stability analysis.

A dramatically different feature of higher dimensional black holes is their ability to take on
nonspherical topologies \cite{Galloway2006}. In $D=5$, the initial discovery of the Emparan-Reall (singly spinning) black ring \cite{emparan2002rotating} with horizon topology $S^1\times S^2$ showed that the no hair theorem definitively fails. There is hope, however, that this theorem could be revived with the additional hypothesis of stability. Namely, it is conjectured that there is a unique \textit{stable} stationary vacuum black hole determined by its mass and angular momenta, and that this solution is a (slowly rotating in $D>5$) Myers-Perry black hole. This suggests that black rings, and the more complicated black lenses \cite{kunduri2014supersymmetric,Tomizawa:2016kjh}, are unstable. Indeed, recently there has been much numerical work all of which gives strong evidence that the family of black ring solutions is unstable \cite{Reall0,Santos:2015iua,Figueras:2015hkb}. It is therefore natural to assume that a mass-angular momentum inequality for black ring (and black lens) spacetimes is not possible. Surprisingly this turns out not to be the case, and it is the purpose of this letter to establish an inequality relating mass and angular momentum for black ring spacetimes, namely
\begin{equation}\label{BRineq}
m^3\geq \frac{27\pi}{4} |\mathcal{J}_{2}|\left|\mathcal{J}_{1}-\mathcal{J}_{2}\right|
\end{equation}
where $\mathcal{J}_1$ and $\mathcal{J}_2$ are the angular momenta associated with $S^1$ and $S^2$, respectively. This inequality is derived in the same fashion as the previous two inequalities, with the final state being the doubly spinning Pomeransky-Sen'kov (PS) black ring \cite{Pomeransky2006}. The validity of this inequality then offers indirect evidence for WCC and FSC, in the setting of $D=5$ black ring spacetimes. In particular, it provides implicit support for the possible nonlinear stability of the PS black rings in contrast to recent developments.

Interpreted with Newtonian considerations, \eqref{BRineq} states that the gravitational binding force of the black hole ($\sim m^2/r^3$) dominates the centripetal repulsive forces ($\sim \mathcal{J}^2/mr^3$) 
to prevent the system from flying apart.  This simple picture does not, however,  differentiate between \eqref{a2} and \eqref{BRineq}.   In addition, the regularity condition $|\mathcal{J}_1|\geq|\mathcal{J}_2|$ will be imposed, which inhibits the black ring from collapsing. We also point out that \eqref{BRineq} is sharp in the sense that it admits a
rigidity statement, identifying the extreme PS black ring as the only spacetime which
saturates the inequality. This then naturally provides a variational characterization of the extreme PS black rings as possessing the absolute minimal mass with fixed angular momentum
among black ring spacetimes.

In order to establish these claims, consider a maximal initial data set $(M^4, g, k)$ consisting of a complete Riemannian 4-manifold with metric $g$ and extrinsic curvature $k$ satisfying the
constraint equations
\begin{equation}\label{eq.1}
16\pi\mu = R-|k|^{2},\quad\quad 8\pi J = \operatorname{div}k.
\end{equation}
Here $\mu$ and $J$ are energy and momentum density of the matter fields, and $R$ denotes the
scalar curvature of $g$. The data has one designated asymptotically flat end from which the ADM mass $m$ arises, and a second end which is either Kaluza-Klein-asymptotically flat (KK-AF) or asymptotically cylindrical (AC).  In both cases the topology of this second end is $(0,\infty) \times S^1 \times S^2$.  In addition it is assumed that the data set is bi-axisymmetric, that is
\begin{equation}\label{eq.2}
	\mathfrak{L}_{\eta_{(i)}}g=\mathfrak{L}_{\eta_{(i)}}k
	=\mathfrak{L}_{\eta_{(i)}}\mu=\mathfrak{L}_{\eta_{(i)}}J=0,
\end{equation}
where $\eta_{(i)}$, $i=1,2$ are the two Killing field generators associated with the $U(1)^2$-action, and $\mathfrak{L}_{\eta_{(i)}}$ denotes Lie differentiation.
This is then referred to as a \emph{black ring initial data set} if $M^4$ is diffeomorphic to $\R^4\#(S^2\times D^2)$, where $D^2$ is the open unit disk. The orbit space $M^4/U(1)^2$ is diffeomorphic to the right-half plane $\{ (\rho,z) |\rho \geq 0\}$ \cite{Hollands2008} such that the $z$-axis $\Gamma$ is divided into intervals serving as axes of rotation for the symmetry generators. In particular $\Gamma=I_1\cup I_2\cup I_3$ with $I_1=(a,\infty)$, $I_2=(0,a)$, and $I_3=(-\infty,0)$ where the Killing vectors $\eta_{(1)}$ and $\eta_{(2)}$ vanish on $I_1$ and $I_2\cup I_3$, respectively. The point $(0,a)$ serves as a `corner' where two rotation axes meet, and $(0,0)$ represents the second end with ring type topology.

This class of data is motivated as follows. According to \cite{alaee2013notes,chrusciel2010global}, the maximal constant time slices in the domain of outer communication of the stationary black ring family are diffeomorphic to $\R^4\#(S^2\times D^2)$. Naturally, the extreme black ring data is complete with two ends, one asymptotically flat and the other AC; their orbit space is as above.
For the non-extreme black ring, however, the orbit space rod structure has an additional interval $H$ representing the $S^1\times S^2$ boundary horizon, that is $\Gamma=I'_1\cup I'_2 \cup H \cup I'_3$ with $I'_1=(c,\infty)$, $I'_2=(b,c)$, $H=(-b,b)$, and $I'_3=(-\infty,-b)$. To obtain a complete manifold, it is standard to double the data by reflecting across the horizon, to obtain two isometric copies attached along the horizon.  This `doubling' procedure results in a manifold with two asymptotically flat ends, and has topology $\R^4\#(S^2\times S^2) \# \R^4$.  Thus, in marked contrast with the familiar case of spherical horizons in which doubling produces $\mathbb{R}\times S^3$, the complete (doubled) non-extreme black rings have \textit{different} topology than their extreme counterparts and thus do not fall into the category of black ring initial data. Nonetheless the class of data considered here does include a suitable completion of the non-extreme stationary black ring. In particular, one way to show this is to appropriately modify certain global existence results \cite{BartnikQS} to construct a solution of the vacuum
constraints on a second KK-AF end, which may then be glued at the horizon to the domain of outer communication.

The metric $g$ is determined by two functions $U$ and $\alpha$, a symmetric $2\times 2$ matrix $\lambda_{ij}$ with $\det\lambda=\rho^2$, and a pair of gauge fields $A^{(i)}$, $i=1,2$ on the orbit space, all of which are specified using Brill's ansatz \cite{brill1959positive}. Let $(\rho,z,\phi^{1},\phi^{2})$ be global cylindrical (Brill) coordinates where $\phi^i\in[0,2\pi]$ correspond to the rotational Killing
directions, that is $\eta_{(i)}=\partial_{\phi^i}$. Then the metric is given as follows with all coefficients independent of rotational coordinates
\begin{equation}\label{eq.6}
\begin{split}
g=&\frac{e^{2U+2\alpha}}{2\sqrt{\rho^2+z^2}}\left(d\rho^2+d z^2\right)\\
&\qquad+e^{2U}\lambda_{ij}\left(d\phi^i+A^{(i)}\right)\left(d\phi^j+A^{(j)}\right).
\end{split}
\end{equation}
Appropriate asymptotics for the coefficients in the designated AF end and AC end may be found in  \cite{alaee2016proof}.  Along the KK-AF end as $r^2=2\sqrt{\rho^2+z^2}\rightarrow 0$, the primary coefficients exhibit the following behavior: $U\sim -2\log r$, $\alpha\sim-\log r$, and $\lambda_{ij}\sim \mathrm{diag}(\rho^2/r^4,r^4)$.

The relevant part of the second fundamental form is characterized by two potentials which
encode the angular momentum. Consider the 1-form $\mathcal{P}_{(i)}=2\star \left(\operatorname{div}k(\eta_{(i)})\wedge\eta_{(1)}\wedge\eta_{(2)}\right)$ on $M^{4}$, where  $\star$ is the Hodge dual. A computation utilizing the momentum constraint $J(\eta_{(i)})=0$, $i=1,2$, and the fact that $\eta_{(i)}$ is a Killing field show that $d\mathcal{P}_{(i)}=0$ \cite{alaeekhurikunduri}. Since $M^{4}$ is simply connected, twist potentials exist globally and satisfy $d\zeta^{i}=\mathcal{P}_{(i)}$.
It follows from the definition of $\mathcal{P}_{(i)}$ that these potentials are constant on the positive ($\Gamma_{+}$) and negative ($\Gamma_{-}$) $z$-axis. Moreover the ADM angular momenta arises as the difference between these constants
\begin{equation}\label{eq.16}
		\mathcal{J}_{i}=\frac{\pi}{4}(\zeta^{i}|_{\Gamma_{-}}-\zeta^{i}|_{\Gamma_{+}}).
\end{equation}

In $D=5$ the stationary bi-axisymmetric vacuum Einstein equations reduce to a sigma model \cite{maison1979ehlers} with domain $\mathbb{R}^3\setminus\Gamma$ and target space $SL(3,\mathbb{R})/SO(3)\cong\mathbb{R}^5$. For general black ring initial data (off shell) the
fields which relate to the latent sigma model structure are $\Psi=(U,\lambda_{ij},\zeta^i)$.
Assuming nonnegative energy density $\mu\geq 0$, an analysis of the Einstein-Hilbert action yields a lower bound \cite{alaee2016proof} for the ADM mass in terms of these variables
\begin{equation}\label{eq.18}
m\geq\mathcal{M}(\Psi),
\end{equation}
where the \textit{mass functional} is given by
\begin{equation}\label{eq.19}
\begin{split}
	&\mathcal{M}(\Psi)
=\frac{1}{8}\int_{\mathbb{R}^{3}}
\left(\frac{e^{-6U}}{2\rho^{2}}\sum_{i,j=1}^2\lambda^{ij}\delta_{3}(\nabla\zeta^{i},\nabla\zeta^{j})\right.\\
&\left.+6|\nabla U|^{2}-\frac{\det\nabla\lambda}{2\rho^{2}} \right)dx+\frac{\pi}{2}\sum_{i=1}^3\int_{I_i}\alpha(0,z)dz.
\end{split}
\end{equation}
Here $\delta_{3}=d\rho^{2}+d z^{2}+\rho^{2} d\phi^{2}$ is a flat metric
on an auxiliary $\mathbb{R}^{3}$, and $\nabla$ and $dx$ are the connection and volume form with respect to $\delta_3$.

Unfortunately \eqref{eq.19} is not manifestly nonnegative and does not exhibit the underlying
sigma model structure. This, however, may be rectified by an appropriate change of variables.
Observe that since $\det\lambda=\rho^{2}$, there are only two independent functions contained in the matrix $\lambda$. Thus, guided by the geometry on the axis, appropriate new variables $(V,W)$ may be constructed and implicitly defined through the relations
\begin{equation}\label{eq.20}
\begin{split}
	\lambda_{11}=f^a_- e^{V}\cosh W,& \quad\quad
\lambda_{22}=f^a_+ e^{-V}\cosh W,\\
\lambda_{12}&=\rho\sinh W,
\end{split}
\end{equation}
where $f^a_{\pm}=\sqrt{\rho^2+(z-a)^2}\pm(z-a)$. Then using the fact that conical singularities
are absent on the axes $I_{i}$, which is equivalent to
\begin{equation}\label{eq.21}
2\alpha (0,z)=(-1)^{\frac{i(i-1)}{2}}{V}(0,z)+\log\left(\frac{|z|}{|z-a|}\right),
\end{equation}
a computation shows that
\begin{equation}\label{eq.23}
	\begin{split}
		\mathcal{M}(\Psi)=&\frac{1}{16}\int_{\mathbb{R}^{3}}12|\nabla U|^{2}+|\nabla V|^{2}+|\nabla W|^{2}dx\\
		&+\frac{1}{16}\int_{\mathbb{R}^{3}}\sinh^{2}W|\nabla (V+h_{2})|^{2} dx\\
		&+\frac{1}{16}\int_{\mathbb{R}^{3}}
		e^{-6h_{1}-6U+h_{2}+V}\cosh W
		\\
		&\qquad \times\left|e^{-h_{2}-V}\tanh W\nabla \zeta^{1}-\nabla \zeta^{2}\right|^{2}dx\\
		&+\frac{1}{16}\int_{\mathbb{R}^{3}}
		\frac{e^{-6h_{1}-6U-h_{2}-V}}{\cosh W}|\nabla \zeta^{1}|^{2}dx,
	\end{split}
\end{equation}
where the functions $h_{1}=\frac{1}{2}\log\rho$ and $h_{2}=\frac{1}{2}\log\left(\frac{f^a_-}{
f^a_+}\right)$ are harmonic on  $\mathbb{R}^{3}\setminus\Gamma$ with respect to $\delta_{3}$.

The expression \eqref{eq.23} clearly identifies the mass lower bound as the reduced energy for
a sigma model, and provides a version of the positive mass theorem for black ring initial data.
Setting $u=U+h_{1}$, $v=V+h_{2}$, and $w=W$ and using integration by parts shows that the reduced energy  $\mathcal{M}$ is up to boundary terms the harmonic energy $E(\tilde{\Psi})$ of a singular map $\tilde{\Psi}=(u,v,w,\zeta^{1},\zeta^{2}):\mathbb{R}^{3}\setminus\Gamma\rightarrow
SL(3,\mathbb{R})/SO(3)$ in which the symmetric space target is naturally endowed with a left invariant metric of nonpositive curvature \cite{maison1979ehlers}. In particular, for compact domains $\Omega\subset \mathbb{R}^3\setminus \Gamma$ the following relation holds
\begin{equation}\label{eq.26}
\begin{split}
16\mathcal{M}_{\Omega}(\Psi)=&E_{\Omega}(\tilde{\Psi})
-12\int_{\partial\Omega}
(h_{1}+2U)\partial_{\nu}h_{1}\\
&-\int_{\partial\Omega}(h_{2}+2V)\partial_{\nu}h_{2},
\end{split}
\end{equation}
where $\nu$ denotes the unit outer normal to the boundary $\partial\Omega$. This shows that
the two functionals have the same critical points. Moreover, since the extreme PS black ring solves the stationary bi-axisymmetry vacuum Einstein equations, its associated map $\Psi_0
=(U_0,V_0,W_0,\zeta^{1}_0,\zeta^{2}_0)$ is a critical point. In fact it is a global minimum as expressed by the following energy gap theorem.

\begin{theorem}\label{infimum}
Among all black ring initial data with fixed angular momenta, the extreme PS black ring achieves the minimum reduced energy. More precisely, given $\Psi$ arising from black ring initial data, and the extreme PS black ring map $\Psi_0$ with
$\zeta^{i}|_{\Gamma}=\zeta^{i}_{0}|_{\Gamma}$, $i=1,2$, we have
\begin{equation}\label{53}
		\mathcal{M}(\Psi)-\mathcal{M}(\Psi_{0})
		\geq C\left(\int_{\mathbb{R}^{3}}
		\operatorname{dist}^{6}(\Psi,\Psi_{0})dx
		\right)^{\frac{1}{3}}
\end{equation}
for some universal constant $C>0$.
\end{theorem}

The proof is based on the convexity of harmonic energy along geodesic deformations
in nonpositively curved target spaces \cite{schoen2013convexity}. In the current setting, however, this cannot be applied directly since the maps $\tilde{\Psi}$, at which the energy $E$ is evaluated, are singular on the axes. Thus, it must be shown that convexity is inherited by the finite reduced energy, and this is accomplished with a cut-and-paste procedure. Let $\delta,\varepsilon>0$ be
small parameters and define sets $\Omega_{\delta,\varepsilon}=\{\delta< r<2/\delta;
\rho>\varepsilon\}$ and $\mathcal{A}_{\delta,\varepsilon}=B_{2/\delta}\setminus
\Omega_{\delta,\varepsilon}$, where $B_{2/\delta}\subset\mathbb{R}^3$ is the ball of radius $2/\delta$ centered at the origin. Smooth cut-off functions may then be used to construct cut-and-paste data $\Psi_{\delta,\varepsilon}$ which agrees with the PS black ring harmonic map $\Psi_0$ near the axes and at spatial infinity, and agrees with $\Psi$ elsewhere. More precisely, the support of the difference of component functions satisfies
\begin{equation*}\label{eq.27}
\begin{split}
&\operatorname{supp}(U_{\delta,\varepsilon}-U_{0})\subset B_{2/\delta},\\
&\operatorname{supp}(V_{\delta,\varepsilon}-V_{0},W_{\delta,\varepsilon}-W_{0},
\zeta^{1}_{\delta,\varepsilon}-\zeta^{1}_{0},\zeta^{2}_{\delta,\varepsilon}-\zeta^{2}_{0})\subset \Omega_{\delta,\varepsilon}.
\end{split}
\end{equation*}
As in \cite{alaee2016proof}, it can be shown that the change in reduced energy from this cut-and-paste construction can be made arbitrarily small, that is
\begin{equation}\label{106}
	\lim_{\delta\rightarrow 0}\lim_{\varepsilon\rightarrow 0}
	\mathcal{M}(\Psi_{\delta,\varepsilon})=\mathcal{M}(\Psi).
\end{equation}
Next let $\tilde{\Psi}_{t}$, $t\in[0,1]$ be a geodesic in $SL(3,\mathbb{R})/SO(3)$ which connects
$\tilde{\Psi}_{\delta,\varepsilon}$ and $\tilde{\Psi}_{0}$. By the construction above $\Psi_{t}\equiv\Psi_{0}$ outside $B_{2/\delta}$ and on a neighborhood of $\mathcal{A}_{\delta,\varepsilon}$, so that in these regions the geodesic is linear in the first two components, $U_{t}=U_{0}+t(U_{\delta,\varepsilon}-U_{0})$ and $V_{t}=V_{0}$. Then since $E$ is convex along geodesics, by using relation \eqref{eq.26} and linearity of $U_t$ and $V_t$ to handle the boundary terms, we have
\begin{equation}\label{eq.28}
	\frac{d^{2}}{dt^{2}}\mathcal{M}(\Psi_{t})
	\geq 2\int_{\mathbb{R}^{3}}|\nabla\operatorname{dist}(\Psi,\Psi_{0})|^{2}dx.
\end{equation}
In addition, since $\Psi_{0}$ is a critical point of $\mathcal{M}$, by integrating \eqref{eq.28} twice and applying a Sobolev inequality, the proof is complete. We arrive at the main result.

\begin{theorem}\label{TheoremI}
Let $(M^4,g,k)$ be a black ring initial data set with nonnegative energy density $\mu\geq 0$,
and zero momentum density in the direction of rotation $J(\eta_{(i)})=0$, $i=1,2$. If the circular angular momentum dominates the spherical, that is $|\mathcal{J}_1| \geq |\mathcal{J}_2|$, then the mass-angular momentum inequality \eqref{BRineq} holds.
Moreover if $|\mathcal{J}_{1}|>|\mathcal{J}_{2}|>0$, then \eqref{BRineq} is saturated if and only if the data arise from the canonical slice of an extreme Pomeransky-Sen'kov black ring spacetime.
\end{theorem}

The main ideas in the proof are the following. By choosing appropriate orientations for rotation, it may be assumed without loss of generality that both angular momenta are positive. The regularity condition $\mathcal{J}_1>\mathcal{J}_2>0$ ensures that the extreme PS black ring spacetime with these angular momenta is nonsingular. Let $\tilde{\Psi}_0$ be the harmonic map associated with this spacetime, then a calculation shows that the mass is given by
\begin{equation}\label{128}
	m_0=\mathcal{M}(\Psi_{0})=\left( \frac{27\pi}{4}\mathcal{J}_{2}(\mathcal{J}_{1}-\mathcal{J}_{2})\right)^{\frac{1}{3}} ;
\end{equation} see the $(\mathcal{J}_1, \mathcal{J}_2)$ phase diagram at fixed $m$ given in \cite{BRedGH:2012} for a more general statement.
The desired mass-angular momentum inequality \eqref{BRineq} now follows by combining \eqref{eq.18}, \eqref{128}, and Theorem \ref{infimum}. Moreover, if equality is achieved in \eqref{BRineq} then according to Theorem \ref{infimum} we have that $\Psi=\Psi_0$, and from here analogous arguments to those in \cite{alaee2016proof} imply that the data $(M^4,g,k)$ coincides with the canonical slice of the Pomeransky-Sen'kov black hole. If it is only assumed that $|\mathcal{J}_1|\geq|\mathcal{J}_2|$, then by perturbing the initial data to achieve a strict regularity condition, the above arguments apply and yield the mass-angular momentum inequality for the perturbation. By taking a limit we find that the original data must also satisfy the inequality.

We observe that for the subset of data with $\mathcal{J}_2=0$, the inequality \eqref{BRineq} reduces to the positive mass theorem.  The canonical example of such data is that corresponding to the Emparan-Reall black ring. There is no rigidity statement in this case, as extreme rings must have $\mathcal{J}_2 > 0$.

These results demonstrate another qualitative feature of black rings differentiating them from spherical black holes. For the latter, at fixed mass it is clear from \eqref{a2} that there is an upper bound on the total magnitude of the angular momenta.  In contrast, \eqref{BRineq} implies that only a certain combination of spins $\mathcal{J}_i$ is bounded.  In particular, spin along the $S^1$ direction of the ring can become arbitrarily large. This was known to be true in the stationary case, but our result shows that it is a characteristic of black rings in the dynamical regime as well. We expect that the methods developed here to treat the ring case will lead 
to analogous results for black holes of other exotic topologies.

In summary, we have proven a highly nontrivial relation between the mass and angular momentum of
spacetimes admitting black holes of ring type. This inequality is intimately tied with and derived from the FSC and WCC conjectures. The FSC and WCC underlie our fundamental understanding of gravitational collapse and formation of black holes. Indeed, the observational data from gravitational waves emitted during binary black hole mergers \cite{Abbott:2016blz} corroborates the expectation that the endpoint is a stationary Kerr black hole.  A failure of WCC in general relativity implies a serious breakdown of the theory in the sense that predictive power is lost in the presence of naked singularities.  The geometric inequality \eqref{BRineq} provides rigorous evidence in support of these two conjectures in the setting of the dynamical evolution of black rings.  This result is particularly striking because recent numerical work, focusing on dynamical black rings \cite{Figueras:2015hkb} and `ultraspinning' black holes \cite{Figueras:2017zwa}, suggests violations of WCC occur in $D \geq 5$.  We anticipate future investigations should resolve the tension between these two sets of results.


{\bf Acknowledgments.} A. Alaee acknowledges the support of a PIMS Postdoctoral Fellowship. M. Khuri acknowledges the support of NSF Grant DMS-1308753. H. Kunduri acknowledges the support of NSERC Grant 418537-2012.

\bibliographystyle{apsrev4-1}
\bibliography{masterfileBR}
\end{document}